\documentclass[sigconf,noacm]{acmart}

\usepackage{tcolorbox}
\usepackage{multirow}
\usepackage{tikz}
\usepackage{amsmath}
\usepackage{wrapfig}
\usepackage{float}
\usepackage{pifont}
\usepackage{wasysym}
\usepackage{fontawesome}
\usepackage{color, colortbl}
\usepackage{arydshln}
\usepackage{url}
\usepackage{tcolorbox}
\usepackage{listings} 

\usepackage[linesnumbered,vlined]{algorithm2e}
\usepackage{graphicx} 
\usepackage{makecell}
\usepackage{enumitem}
\usepackage{subcaption}
\usepackage{changepage}

\renewcommand\footnotetextcopyrightpermission[1]{} 
\definecolor{nickgreen}{HTML}{007007}
\definecolor{dgreen}{HTML}{03A60D}
\definecolor{dred}{HTML}{A6030D}
\definecolor{darkbrown}{rgb}{1, 0.55, 0}

\definecolor{Gray}{gray}{0.9}
\definecolor{LightCyan}{rgb}{0.88,1,1}
\definecolor{Color001}{RGB}{148,0,85}
\definecolor{Color002}{RGB}{0,120,120}
\definecolor{Color003}{RGB}{199,232,172}

\newcommand{\gc}{\cellcolor{Gray}}

\newcommand{\circI}{{\small\Circle}}
\newcommand{\circII}{{\small\LEFTcircle}}
\newcommand{\circIII}{{\small\CIRCLE}}

\AtBeginDocument{%
  }

\setcopyright{acmlicensed}
\copyrightyear{2018}
\acmYear{2018}
\acmDOI{XXXXXXX.XXXXXXX}

\acmConference[]{}{}

\begin{document}

\title{SoK: Self-Sovereign Digital Identities}

\author{Sushanth Ambati}
\email{ambati23@rowan.edu}
\orcid{0009-0003-9209-9094}
\affiliation{%
  \institution{Rowan University}
  \city{Glassboro}
  \state{New Jersey}
  \country{USA}
}

\author{Kainat Adeel}
\email{adeelk57@students.rowan.edu}
\affiliation{%
  \institution{Rowan University}
  \city{Glassboro}
  \state{New Jersey}
  \country{USA}
}

\author{Jack Myers}
\email{myersjac@rowan.edu}
\affiliation{%
  \institution{Rowan University}
  \city{Glassboro}
  \state{New Jersey}
  \country{USA}
}

\author{Nikolay Ivanov}
\orcid{0000-0002-2325-2847}
\email{ivanov@rowan.edu}
\affiliation{%
  \institution{Rowan University}
  \city{Glassboro}
  \state{New Jersey}
  \country{USA}
}

\renewcommand{\shortauthors}{}

\begin{abstract}
Self-Sovereign Digital Identity (SSDI) enables individuals to control their own identity assertions and data, rather than relying on centralized or federated systems prone to large-scale data breaches. By eliminating centralized databases maintained by service providers and identity brokers, SSDIs offer enhanced security and privacy. However, adoption remains slow, and research in this area lacks systematization and uniformity. To address these gaps, we present a comprehensive systematization of knowledge on self-sovereign digital identities, with a primary focus on identifying the challenges that impede real-world adoption. We survey 80 academic and non-academic sources and identify six major challenges: (i) binding a single identity to one individual or organization, (ii) the absence of mature cryptographic and communication protocols, (iii) significant usability barriers, (iv) regulatory and oversight gaps, (v) bootstrapping to critical-mass adoption, and (vi) dependence on a permissionless, decentralized, yet singular infrastructure that may expose unforeseen vulnerabilities over time. We then analyze 47 scientific publications and find that the vast majority focus on blockchain-based solutions rather than generalized SSDI architectures. Additionally, we catalog 12 real-world, production-grade SSDI applications. Our evaluation of these solutions reveals that self-sovereignty is, in practice, a spectrum rather than a binary property. Finally, we explore the frontiers of SSDI by identifying major trends, open problems, and opportunities for future research. We hope this systematization will help advance the shift from centralized to self-sovereign digital identities in a disciplined and impactful way.
\end{abstract}



\maketitle

\section{Introduction}\label{sec:introduction}
\begin{figure*}[htb]
    \centering
    \includegraphics[width=\linewidth]{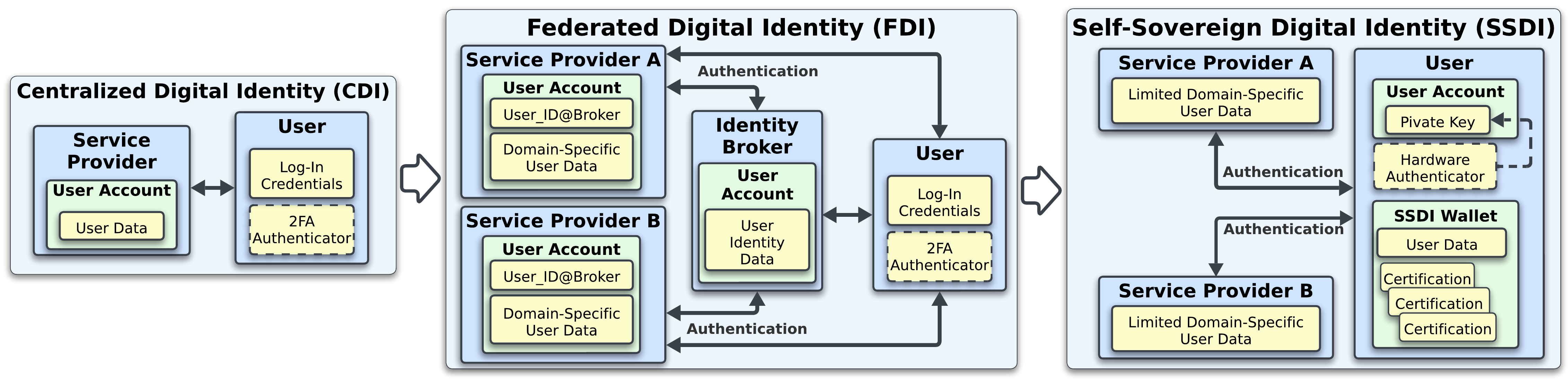}
    \caption{\textbf{Evolution of digital identity paradigms:} Centralized Digital Identity (CDI) is based on the premise that each service provider maintains and stores accounts of its users. Federated Digital Identity (FDI) outsources authentication to an Identity Broker, such as Google, Facebook, Apple, or LinkedIn. Self-Sovereign Digital Identity (SSDI) is the latest paradigm in which user authentication keys and data are stored by the user and provided on demand in the form of digital certifications.}
    \label{fig:di-evo}
\end{figure*}

Digital identity is the foundation upon which virtually every online interaction is built. From logging into a social media platform to accessing government services, the ability to reliably establish ``who someone is'' in digital space underpins authentication, authorization, and access control~\cite{bertino2010identity, torres2012survey}. Historically, digital identity management has followed a trajectory from siloed, centralized systems toward increasingly user-centric paradigms, as illustrated in Fig.~\ref{fig:di-evo}. Today, the prevailing model is the \emph{Federated Digital Identity} (FDI), in which an \emph{Identity Broker} such as Google, Facebook, or Apple mediates authentication on behalf of many service providers~\cite{maler2008venn, grassi2015privacy}. While FDI reduces password fatigue and simplifies account management, it concentrates enormous volumes of personal data in a small number of corporations, creating attractive targets for both external attackers and internal misuse~\cite{hill202115, guardian2018cambridge, guardian2018zuckerberg}.

The consequences of this concentration are not hypothetical. Major data breaches have affected billions of user records~\cite{hammouchi2019digging, ayyagari2012exploratory}, and incidents such as the Cambridge Analytica scandal demonstrated that even lawfully collected identity data can be repurposed in ways that undermine democratic processes~\cite{hinds2020wouldn}. In response, a growing body of research and development has coalesced around the concept of \emph{Self-Sovereign Digital Identity} (SSDI), a paradigm in which individuals and organizations hold, control, and selectively disclose their own identity assertions without dependence on any single intermediary~\cite{allen2016path, preukschat2021self, der2017self}. The conceptual appeal of SSDI is substantial: users gain genuine data ownership, service providers receive only the minimum information they need, and no central database exists to be breached.

Despite this promise, real-world adoption of SSDI remains remarkably limited. The gap between vision and practice can be attributed to a combination of technical, social, and regulatory challenges that the research community has not yet systematically catalogued or prioritized. Existing surveys tend to focus on specific technical layers (for instance, Decentralized Identifiers (DIDs)~\cite{w3c2021didreferences, w3c_did_core_v1_1} or Verifiable Credentials (VCs)~\cite{sporny2019vcmodel, sporny2019vcprimer}), or on a particular application domain such as healthcare~\cite{houtan2020survey} or IoT~\cite{bouras2021lightweight, mahalle2022identity}. What is missing is a holistic, challenge-centric systematization that cuts across layers and domains, connects technical obstacles to socioeconomic realities, and evaluates the state of both scientific literature and production deployments.

This paper fills that gap. We present a comprehensive \emph{Systematization of Knowledge} (SoK) on Self-Sovereign Digital Identities. Our methodology proceeds in five steps (see Fig.~\ref{fig:steps}). First, we analyze existing surveys to establish the current state of understanding (\S\ref{sec:prior}). Second, drawing on 80 sources that include research papers, industry projects, company reports, and cypherpunk initiatives (i.e., grassroots privacy-advocacy projects), we distill six major challenges that impede SSDI adoption (\S\ref{sec:challenges}). Third, we conduct a structured analysis of 47 scientific publications and find that the overwhelming majority equate SSDI with blockchain, leaving generalized architectures underexplored (\S\ref{sec:scientific}). Fourth, we catalog 12 production-deployed SSDI applications and evaluate where each falls on the spectrum of self-sovereignty (\S\ref{sec:realworld}). Fifth, we outline the frontiers of SSDI research and practice (\S\ref{sec:frontiers}).

Specifically, our main contributions are:
\begin{enumerate}
    \item \textbf{Challenge taxonomy.} We identify and thoroughly analyze six fundamental challenges: identity binding, protocol maturity, usability, oversight, critical-mass adoption, and single point of trust infrastructure, which collectively explain why SSDI has not achieved widespread deployment.
    \item \textbf{Literature systematization.} We provide a structured analysis of 47 scientific publications, revealing a strong blockchain bias and identifying underexplored research directions.
    \item \textbf{Real-world evaluation.} We catalog and assess 12 production-grade SSDI systems, demonstrating that self-sovereignty is a spectrum rather than a binary property.
    \item \textbf{Research roadmap.} We articulate open problems and emerging trends, from zero-knowledge proofs and homomorphic encryption to regulatory harmonization and Web3 integration, collectively establishing a foundation for future work.
\end{enumerate}

\section{Methodology}\label{sec:methodology}
\begin{figure*}
    \centering
    \includegraphics[width=\linewidth]{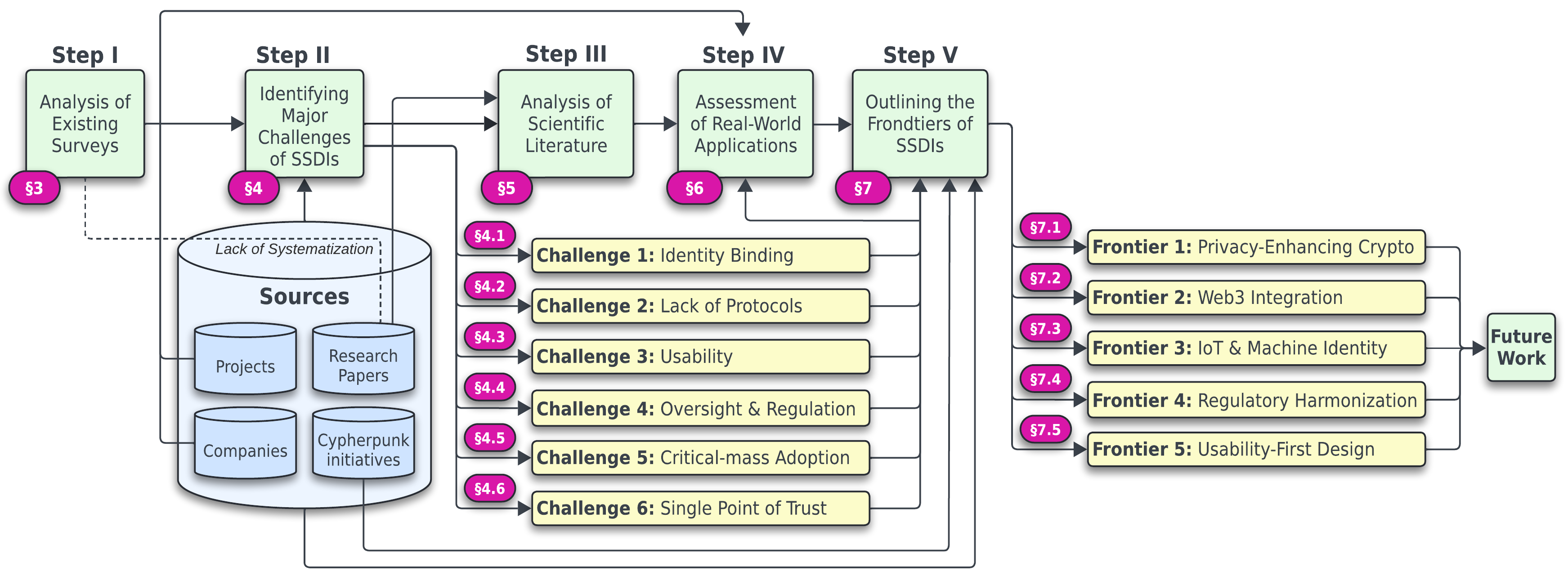}
    \caption{Five-step methodology of this study. We first analyze existing surveys, revealing a near-complete absence of existing systematization. Then we identify six major challenges impeding the adoption of SSDIs. After that, we analyze 47 research papers and observe disproportionate focus on certain areas and challenges. Next, we survey real-world applications of SSDIs, confirming the ``sovereignty washing'' phenomenon and the lack of production deployment. Finally, we identify five major frontiers of SSDI research and outline directions for future work.}
    \label{fig:steps}
\end{figure*}

Our systematization follows a five-step methodology, illustrated in Fig.~\ref{fig:steps}, designed to capture both the breadth and depth of the SSDI landscape.

\textbf{Step~I: Analysis of Existing Surveys (\S\ref{sec:prior}).}
We began by identifying and critically reviewing prior surveys on self-sovereign and decentralized identity. We searched IEEE Xplore, ACM Digital Library, Springer, Elsevier, and Google Scholar using the queries ``self-sovereign identity survey,'' ``decentralized identity survey,'' and ``SSI systematization.'' We identified eight substantive prior surveys published between 2018 and 2023. For each, we recorded its scope, methodology, number of sources analyzed, and whether it addressed challenges beyond the purely technical. This step established the baseline of existing knowledge and confirmed the lack of a challenge-centric systematization.

\textbf{Step~II: Identifying Major Challenges (\S\ref{sec:challenges}).}
To identify the challenges that impede SSDI adoption, we drew from a deliberately broad range of sources. Our source pool comprises 80 sources spanning four categories: (a)~active SSDI \emph{projects} (e.g., Sovrin, uPort, Jolocom), (b)~peer-reviewed \emph{research papers}, (c)~\emph{company} and standards-body reports (e.g., W3C, DIACC, World Economic Forum), and (d)~grassroots privacy-advocacy \emph{initiatives} and manifestos that articulate the philosophical underpinnings of self-sovereignty~\cite{hughes1993cypherpunk, jarvis2022cypherpunk, kaleta2025legacy}. We systematically catalogued the challenges mentioned in each source and iteratively consolidated them until six stable, mutually distinct challenge categories emerged. Importantly, we find that no single prior work recognizes all six challenges simultaneously, confirming the lack of systematization that motivates this paper.

\textbf{Step~III: Analysis of Scientific Literature (\S\ref{sec:scientific}).}
We conducted a structured literature review of 47 scientific publications that propose, design, or evaluate SSDI systems. Inclusion criteria required that a work (i)~present a concrete technical contribution (architecture, protocol, or system) rather than solely a position paper, and (ii)~explicitly target self-sovereign or fully decentralized identity. For each included paper we recorded the underlying trust infrastructure (blockchain, peer-to-peer, other), the cryptographic primitives employed, the identity lifecycle phases addressed (creation, authentication, revocation, recovery), and whether user studies or formal security proofs were provided.

\textbf{Step~IV: Assessment of Real-World Applications (\S\ref{sec:realworld}).}
We identified 12 SSDI systems that have been deployed in production or large-scale pilot settings. For each, we evaluated the degree of self-sovereignty along five dimensions derived from the foundational principles of self-sovereign identity~\cite{allen2016path}: (1)~user control over key material, (2)~absence of a permissioned gatekeeper for credential issuance, (3)~portability of credentials across platforms, (4)~selective disclosure support, and (5)~resistance to unilateral revocation by a third party.

\textbf{Step~V: Outlining the Frontiers (\S\ref{sec:frontiers}).}
Finally, we synthesized the findings from Steps~I--IV to identify major trends, open problems, and opportunities for future research and standardization. Specifically, we cross-referenced the gaps identified in Steps~II and~III with the deployment realities observed in Step~IV to surface the most pressing directions for both the research community and practitioners.

\section{Prior Surveys}\label{sec:prior}
Several surveys have addressed self-sovereign and decentralized identity, each with a distinct scope and focus. We summarize the most relevant prior works below, organized thematically, and explain how our systematization differs.

\paragraph{Architectural overviews.}
M\"{u}hle et al.~\cite{muhle2018survey} provide one of the earliest surveys on SSI, identifying essential components such as DIDs, verifiable credentials, and distributed ledgers. Their work offers a valuable architectural overview but does not systematically analyze challenges to adoption or evaluate real-world deployments. Similarly, Soltani et al.~\cite{soltani2021survey} survey the SSI ecosystem with an emphasis on technical building blocks: DID methods, credential formats, and wallet architectures; but their treatment of non-technical barriers such as regulation and usability remains limited in scope.

\paragraph{Foundational and practitioner-oriented works.}
Der et al.~\cite{der2017self} present an early and influential discussion of opportunities and challenges for self-sovereign identity. While their work anticipates several of the challenges we formalize (particularly around key management and legal recognition), it is structured as a position paper rather than a systematic survey and does not evaluate scientific literature or production systems. Preukschat and Reed~\cite{preukschat2021self} authored a comprehensive book on SSI that serves as an excellent introduction for practitioners, covering governance frameworks, trust architectures, and use cases. However, it does not adopt a systematic methodology for identifying challenges or analyzing the research landscape.

\paragraph{Broad and domain-specific surveys.}
Ahmed et al.~\cite{bai2022blockchain} provide a comprehensive survey that covers both blockchain-based identity management and the SSI ecosystem. Their work is notable for its breadth, covering over 100 papers, but its scope is so broad that SSDI-specific challenges receive limited attention within the larger blockchain identity narrative. Houtan et al.~\cite{houtan2020survey} focus specifically on blockchain-based self-sovereign identity in healthcare, providing valuable domain-specific insights but leaving cross-domain challenges unaddressed.

\paragraph{Ecosystem and sociopolitical perspectives.}
Laatikainen et al.~\cite{laatikainen2021self} adopt an ecosystem perspective, examining the benefits and challenges of SSI through the lens of platform economics. Their work highlights the bootstrapping problem of adoption (which we formalize as Challenge~5) but does not systematically analyze the technical literature. Giannopoulou~\cite{Giannopoulou2023Digital} offers a critical sociopolitical analysis of SSI, questioning whether decentralization truly empowers users or merely shifts power to different intermediaries. Their examination of power dynamics in decentralized systems directly informs our discussion of the self-sovereignty spectrum in \S\ref{sec:realworld}.

\paragraph{Positioning of this work.}
In contrast to these works, our systematization is the first to (i)~adopt a challenge-centric taxonomy derived from a deliberately diverse source base spanning academic, industrial, and grassroots origins, (ii)~quantitatively characterize the scientific literature's blockchain bias, (iii)~evaluate production deployments along a spectrum of self-sovereignty, and (iv)~synthesize these analyses into a unified research roadmap.

\section{Identifying Major Challenges}\label{sec:challenges}
Through our analysis of 80 sources spanning academic papers, industry projects, standards bodies, and grassroots privacy-advocacy initiatives, we identify six major challenges that collectively impede the real-world adoption of SSDIs. These challenges are interrelated: progress on one often requires or enables progress on others. Nevertheless, each is conceptually distinct and demands targeted research and engineering effort.

\subsection{Challenge~1: Identity Binding}\label{sec:binding}

The most fundamental challenge in any identity system is binding a digital representation to exactly one real-world entity: an individual, organization, or device. In centralized and federated systems, identity binding is enforced by a trusted authority (a service provider or identity broker) that verifies government-issued documents, biometrics, or other credentials during enrollment~\cite{nist2020digital, gelb2017identification}. SSDI, by definition, seeks to eliminate or minimize reliance on such authorities, creating a tension between self-sovereignty and reliable binding.

The core problem is the \emph{Sybil attack}: without a centralized enrollment authority, what prevents a single actor from creating multiple identities?~\cite{stokkink2021truly} In permissionless blockchain networks, for example, anyone can generate an arbitrary number of key pairs and, consequently, an arbitrary number of DIDs~\cite{antonopoulos2018mastering, wood2014ethereum}. This is an intentional property of permissionless systems, but it means that identity binding must be achieved through additional mechanisms.

Several approaches have been proposed. \emph{Credential-based binding} relies on trusted issuers (governments, universities, employers) to attest that a DID corresponds to a specific real-world entity~\cite{sporny2019vcmodel}. This approach is pragmatic but reintroduces a dependency on centralized authorities for initial attestation. \emph{Web-of-trust binding} allows peers to vouch for one another's identity, distributing trust across a social graph~\cite{allen2016path}. While philosophically aligned with self-sovereignty, web-of-trust models are vulnerable to collusion and suffer from bootstrapping problems in sparse networks. \emph{Biometric binding} ties an identity to a physical characteristic, but raises privacy concerns and is irreversible if compromised~\cite{lyon2008biometrics}.

A related conceptual difficulty concerns the nature of digital identity itself. Floridi~\cite{floridi2011informational} argues that personal identity is fundamentally informational, suggesting that digital identity need not mirror physical identity. This view supports the idea of context-dependent partial identities, where a user presents different facets of their identity to different verifiers. However, this multiplicity complicates binding: if a user legitimately maintains multiple partial identities, distinguishing legitimate use from Sybil behavior becomes significantly harder.

\subsection{Challenge~2: Key Management and Protocols}\label{sec:protocols}

SSDI shifts the locus of control from service providers to users, which necessarily means that users must manage their own cryptographic keys~\cite{preukschat2021self, der2017self}. Key management encompasses generation, storage, backup, recovery, rotation, and revocation, each of which is non-trivial even for expert users~\cite{ferguson2011cryptography}.

Key \emph{generation} in SSDI typically follows the hierarchical deterministic (HD) wallet pattern borrowed from cryptocurrency systems, where a single seed phrase generates a tree of key pairs~\cite{wuille2012bip32, palatinus2013bip}. While this approach simplifies backup (the user need only secure the seed phrase), it creates a catastrophic single point of failure: anyone who obtains the seed phrase gains full control of the user's identity. Mnemonic seed phrases (typically 12 or 24 words) are the dominant backup mechanism, but studies consistently show that users struggle to store them securely~\cite{zaeem2021ssi, albayati2021study}.

Key \emph{recovery} is arguably the most critical unsolved problem. In centralized systems, a forgotten password can be reset via email or customer support. In SSDI, there is no ``forgot password'' button. Social recovery schemes, in which a quorum of trusted contacts can collectively restore access, have been proposed but introduce their own trust assumptions and coordination overhead~\cite{he2018social}. Hardware-backed solutions (secure enclaves, hardware security modules) can protect keys at rest but do not solve the backup problem if the device is lost or destroyed~\cite{ivanov2021ethclipper, suratkar2020cryptocurrency}.

Beyond key management, the protocol landscape for SSDI remains fragmented. The W3C DID specification~\cite{w3c2021didreferences} provides a syntactic framework, but over 100 DID methods have been registered~\cite{w3c2019didregistry}, each with different trust models, resolution mechanisms, and security properties. The DIDComm protocol~\cite{hardman2019didcomm} and Peer DIDs~\cite{hardman2019peerdid} address pairwise communication, but interoperability across DID methods remains limited. The Trust over IP (ToIP) stack~\cite{davie2019trust} proposes a layered architecture analogous to the TCP/IP stack, but widespread adoption of this framework is still nascent. In practice, this fragmentation means that credentials issued in one SSDI ecosystem often cannot be verified in another, undermining the portability that self-sovereignty promises.

\subsection{Challenge~3: Usability}\label{sec:usability}

For any security technology, usability is a precondition for adoption, not a secondary concern~\cite{borcea2006what}. SSDI faces particularly steep usability barriers because it asks users to assume responsibilities previously handled by service providers: managing cryptographic keys, understanding credential flows, and making trust decisions about issuers and verifiers~\cite{zaeem2021ssi}.

Zaeem et al.~\cite{zaeem2021ssi} conducted one of the few empirical usability studies of SSDI solutions. Their findings highlight significant barriers: users found the onboarding process confusing, struggled to understand the relationship between DIDs and verifiable credentials, and expressed anxiety about the consequences of losing their private keys. The mental model mismatch is significant, as most users are accustomed to the ``username-and-password'' paradigm and find the shift to key-based authentication conceptually alien.

Wallet design is a critical usability factor. O'Donnell~\cite{odonnell2019wallets} surveyed the state of digital wallets and found significant variation in design maturity. Some wallets present credentials in a familiar ``card'' metaphor, while others expose raw cryptographic details. Albayati et al.~\cite{albayati2021study} studied the user experience of cryptocurrency wallets (which share architectural similarities with SSDI wallets) and found that poor interface design was a major barrier to adoption, even among technically literate users.

The usability challenge is compounded by the credential verification flow. In a typical SSDI interaction, a verifier requests specific credentials, the user reviews the request, selects which credentials to share, and optionally applies selective disclosure to reveal only the necessary attributes. Each step introduces cognitive load. Selective disclosure, while privacy-enhancing, requires users to understand what information is being revealed and what is being withheld, a distinction that is non-trivial even for privacy-conscious individuals. Zero-knowledge proofs~\cite{sun2021survey, yang2020zero}, which enable a user to prove a property (e.g., ``I am over 18'') without revealing the underlying data (e.g., their date of birth), further amplify this challenge: the concept is powerful but deeply unintuitive for non-experts.

\subsection{Challenge~4: Oversight and Regulation}\label{sec:oversight}

SSDI exists in a regulatory gray zone. Most existing legal frameworks for digital identity assume the presence of a centralized identity provider that can be held accountable for data protection, fraud prevention, and law enforcement cooperation~\cite{nist2020digital, oecd2021digital}. SSDI disrupts this assumption by distributing control to the individual, raising novel questions about liability, compliance, and governance.

The European Union's General Data Protection Regulation (GDPR) illustrates the tension. GDPR grants individuals the ``right to be forgotten''~\cite{herian2020blockchain}, but if identity attestations are anchored on an immutable blockchain, deletion may be technically infeasible. Some SSDI systems address this by storing only hashes or pointers on-chain while keeping personal data off-chain, but the legal sufficiency of this approach remains contested~\cite{forbes2018gdpr}. The EU's recently adopted eIDAS~2.0 regulation represents a significant step toward recognizing decentralized identity, mandating that member states provide citizens with a European Digital Identity Wallet. However, the eIDAS framework still requires \emph{qualified trust service providers}, potentially reintroducing centralization.

In the United States, digital identity regulation is fragmented across federal and state levels~\cite{Brennen2024The, Tsukayama2024State}. There is no federal digital identity law, and states are experimenting with a patchwork of approaches. Age verification legislation has become a flashpoint for debates about privacy and identity~\cite{Tsukayama2024EFF, Buckley2025Age}, with critics arguing that mandated identity verification for online services could undermine the very privacy that SSDI aims to protect~\cite{Alajaji2025VPNs}.

Law enforcement access is another fraught area. Centralized identity providers can comply with subpoenas and court orders by producing user records. In an SSDI system, there may be no single entity that possesses the requested data. While this is a privacy feature from the user's perspective, it creates friction with legitimate investigative needs~\cite{Gullo2021EFF}. Striking the right balance between individual sovereignty and societal oversight is a policy challenge that technology alone cannot resolve.

\subsection{Challenge~5: Critical-Mass Adoption}\label{sec:adoption}

SSDI suffers from a classic \emph{cold-start problem}~\cite{laatikainen2021self}. Users will not invest effort in acquiring and managing self-sovereign credentials unless there are service providers (verifiers) that accept them. Conversely, service providers will not invest in integrating SSDI verification unless a critical mass of users presents self-sovereign credentials. Credential issuers (e.g., governments, universities, employers) occupy a pivotal position: they can bootstrap the ecosystem by issuing verifiable credentials, but they face their own incentive misalignments, since adopting SSDI may reduce their centralized control over identity data.

The World Economic Forum has highlighted that over one billion people worldwide lack any form of officially recognized identity~\cite{wef2021digital, id4d2018}. While this population represents a potential greenfield opportunity for SSDI to leapfrog traditional identity infrastructure in developing regions~\cite{wang2019self, gelb2017identification}, realizing this vision requires not only technical infrastructure but also governance frameworks, digital literacy programs, and sustained institutional commitment~\cite{worldbank2016digitalidentity}.

Network effects are essential. SSDI becomes more valuable as more participants join the ecosystem, but the initial value proposition for early adopters is weak. Government mandates (as in the EU's Digital Identity Wallet initiative) can artificially accelerate adoption, but mandates risk undermining the voluntariness that is central to the self-sovereign ethos. Alternatively, niche communities with strong internal incentives, such as professional credentialing in healthcare or education~\cite{grech2021blockchain}, may serve as beachheads from which SSDI can expand to broader populations.

\subsection{Challenge~6: Single Infrastructure Dependence}\label{sec:singlepoint}

The vast majority of SSDI proposals rely on a single blockchain or distributed ledger as their trust anchor~\cite{soltani2021survey, muhle2018survey, bai2022blockchain}. While individual blockchains are internally decentralized, the ecosystem-level dependence on \emph{one} infrastructure creates a \emph{meta-centralization} risk that is rarely examined in prior work.

If the chosen blockchain suffers a consensus failure, a 51\% attack, a critical smart contract vulnerability~\cite{ivanov2023security, ivanov2021targeting}, or a governance crisis, every identity anchored to that chain is simultaneously at risk. Blockchain scalability limitations~\cite{zhou2020solutions, xie2019survey} may also become a bottleneck as the number of identity operations grows. Stokkink et al.~\cite{stokkink2021truly} explicitly argue that a ``truly self-sovereign'' identity system should not depend on any single ledger and propose a peer-to-peer alternative, but their system has not achieved significant adoption.

The problem is not limited to security. Economic sustainability is also a concern: permissionless blockchains require transaction fees, which means that every identity operation (creating a DID, issuing a credential, revoking a key) has a direct monetary cost. During periods of network congestion, these costs can spike unpredictably, making SSDI prohibitively expensive for low-resource users. Permissioned ledgers (such as Sovrin's Hyperledger Indy network~\cite{sovrin2018whitepaper}) avoid transaction fees but sacrifice permissionlessness, reintroducing a governance authority that controls who may participate as a network node.

A truly resilient SSDI ecosystem would be \emph{ledger-agnostic}, allowing users to anchor their identities on any suitable infrastructure and migrate between infrastructures as needed. Achieving this requires not only technical standards for cross-ledger interoperability but also governance frameworks that can accommodate heterogeneous trust models. The current landscape falls far short of this ideal.

\section{Scientific Literature}\label{sec:scientific}

To characterize the state of SSDI research, we conducted a structured analysis of 47 scientific publications that propose, design, or evaluate self-sovereign or fully decentralized identity systems. Table~\ref{tab:literature} presents the complete classification. Each paper is categorized along five dimensions: \emph{type} (System, Protocol, Framework, Application-domain, or Analysis), \emph{trust infrastructure} (Blockchain, Peer-to-peer, Hybrid, or Agnostic), \emph{advanced cryptographic techniques} employed beyond standard public-key cryptography (zero-knowledge proofs, homomorphic encryption, or CL-signatures), which of the \emph{six challenges} (\S\ref{sec:challenges}) the paper addresses, and the \emph{evaluation methodology} (formal security proof, user study, implementation benchmark, or informal argument only). We organize our discussion along four cross-cutting dimensions.

\renewcommand{\arraystretch}{1.0}
\begin{table*}[htbp]
\centering
\caption{\textbf{Classification of 47 scientific publications on SSDI.} \textbf{Crypto}: PKC = standard public-key only, ZKP = zero-knowledge proofs, HE = homomorphic encryption, CL = CL-signatures. \textbf{C1--C6}: challenges addressed (\circIII). \textbf{Eval. Method}: FP = formal proof, US = user study, BM = benchmark, $-$ = informal only.}
\label{tab:literature}
\setlength{\tabcolsep}{4pt}
\begin{tabular}{r|l|c|c|c|c:c:c:c:c:c|c|}
\toprule

 & & & & & \multicolumn{6}{c|}{\textit{SSDI Challenges}} & \\
\cmidrule(lr){6-11}
\textbf{\#} & \textbf{Academic Work} & \textbf{Type} & \textbf{Infrastructure} & \textbf{Crypto} & \textbf{C1} & \textbf{C2} & \textbf{C3} & \textbf{C4} & \textbf{C5} & \textbf{C6} & \textbf{Eval. Method} \\
\midrule
1  & Liu et al.~\cite{liu2017identity}          & System/Architecture & Blockchain & PKC & \circIII & \circIII &  &  &  &  & BM \\
\gc 2 \gc & \gc Abraham~\cite{abraham2017self}              & \gc Framework & \gc Blockchain & \gc PKC & \gc \circIII & \gc \circIII & \gc & \gc \circIII & \gc & \gc & \gc $-$ \\
3  & Omar \& Basir~\cite{omar2018identity}       & System/Architecture & Blockchain & PKC &  & \circIII &  &  &  &  & BM \\
\gc 4  & \gc Othman \& Callahan~\cite{othman2018horcrux} & \gc Protocol & \gc Blockchain & \gc CL  & \gc \circIII & \gc \circIII &  \gc &  \gc &  \gc &  \gc & \gc BM \\
5  & Dunphy \& Petitcolas~\cite{dunphy2018first} & Analysis & Blockchain & PKC & \circIII & \circIII & \circIII &  &  &  & $-$ \\
\gc 6  & \gc Faber et al.~\cite{faber2019bpdims}         & \gc System/Architecture & \gc Blockchain & \gc PKC & \gc \circIII & \gc \circIII & \gc & \gc \circIII & \gc & \gc & \gc BM \\
7  & Zhao \& Liu~\cite{zhao2019blockchain}       & System/Architecture & Blockchain & PKC & \circIII &\circIII &  &  &  &  & BM \\
\gc 8  \gc & \gc Kuperberg~\cite{kuperberg2019blockchain}     & \gc Analysis & \gc Blockchain & \gc PKC &  \gc & \gc \circIII & \gc & \gc & \gc \circIII & \gc & \gc $-$ \\
9  & Shen et al.~\cite{shen2019efficient}        & Protocol & Blockchain & HE  &  & \circIII &  &  &  &  & FP \\
\gc 10 & \gc Drozdowski et al.~\cite{drozdowski2019application} & \gc Protocol & \gc Blockchain & \gc HE & \gc \circIII & \gc \circIII & \gc & \gc & \gc & \gc & \gc BM \\
11 & Liu et al.~\cite{liu2020blockchain}          & Analysis & Blockchain & PKC & \circIII & \circIII &  &  &  &  & $-$ \\
\gc 12 & \gc Dib \& Toumi~\cite{dib2020decentralized}    & \gc Framework & \gc Blockchain & \gc PKC & \gc & \gc \circIII &  \gc & \gc \circIII & \gc & \gc & \gc $-$ \\
13 & Yang \& Li~\cite{yang2020zero}              & Protocol & Blockchain & ZKP & \circIII & \circIII &  &  &  &  & FP \\
\gc 14 & \gc Naik \& Jenkins~\cite{naik2020self}         & \gc Analysis & \gc Blockchain & \gc PKC & \gc & \gc \circIII & \gc \circIII & \gc \circIII & \gc & \gc & \gc BM \\
15 & Sarier~\cite{sarier2021efficient}             & Protocol & Blockchain & ZKP & \circIII & \circIII &  &  &  &  & FP \\
\gc 16 & \gc Bhattacharya et al.~\cite{bhattacharya2020bindaas} & \gc Application/Domain & \gc Blockchain & \gc PKC & \gc \circIII & \gc \circIII & \gc & \gc \circIII & \gc & \gc & \gc BM \\
17 & Lesavre et al.~\cite{lesavre2020blockchain} & Framework & Blockchain & PKC & \circIII & \circIII &  & \circIII & \circIII &  & $-$ \\
\gc 18 & \gc Chean et al.~\cite{chean2018authentication} & \gc Protocol & \gc Blockchain & \gc HE  & \gc & \gc \circIII & \gc & \gc & \gc & \gc & \gc BM \\
19 & Bandara et al.~\cite{bandara2021ssi}        & System/Architecture & Blockchain & PKC & \circIII & \circIII &  &  &  &  & BM \\
\gc 20 & \gc Nusantoro et al.~\cite{nusantoro2021blockchain} & \gc System/Architecture & \gc Blockchain & \gc PKC & \gc \circIII & \gc \circIII & \gc & \gc & \gc & \gc & \gc BM \\
21 & Bouras et al.~\cite{bouras2021lightweight}  & System/Architecture & Blockchain & PKC &  & \circIII &  &  &  &  & BM \\
\gc 22 & \gc Grech et al.~\cite{grech2021blockchain}     & \gc Application/Domain & \gc Blockchain & \gc PKC & \gc & \gc & \gc \circIII & \gc & \gc \circIII & \gc & \gc $-$ \\
23 & Sedlmeir et al.~\cite{sedlmeir2021digital} & Analysis & Blockchain & PKC &  & \circIII &  & \circIII &  &  & $-$ \\
\gc 24 & \gc Mulaji \& Roodt~\cite{mulaji2021practicality} & \gc Analysis & \gc Blockchain & \gc PKC & \gc & \gc & \gc & \gc & \gc \circIII & \gc & \gc $-$ \\
25 & \v{C}u\v{c}ko \& Turkanovi\'{c}~\cite{cucko2021decentralized} & Analysis & Blockchain & PKC & \circIII & \circIII & \circIII & \circIII & \circIII & \circIII & $-$ \\
\gc 26 & \gc Maram et al.~\cite{maram2021candid}         & \gc System/Architecture & \gc Blockchain & \gc ZKP & \gc \circIII & \gc \circIII & \gc & \gc & \gc & \gc & \gc BM \\
27 & Stockburger et al.~\cite{stockburger2021blockchain} & Application/Domain & Blockchain & CL & \circIII & \circIII & \circIII & \circIII &  &  & US \\
\gc 28 & \gc Venkatraman \& Parvin~\cite{venkatraman2022developing} & \gc System/Architecture & \gc Blockchain & \gc PKC & \gc & \gc \circIII & \gc & \gc & \gc & \gc & \gc BM \\
29 & Shuaib et al.~\cite{shuaib2022land}         & Application/Domain & Blockchain & PKC &  & \circIII &  & \circIII &  &  & BM \\
\gc 30 & \gc Zhuang et al.~\cite{zhuang2022bcppt}        & \gc System/Architecture & \gc Blockchain & \gc ZKP & \gc \circIII & \gc \circIII & \gc & \gc & \gc & \gc & \gc FP \\
31 & Toth \& Anderson-Priddy~\cite{toth2019self}  & Analysis & Blockchain & PKC &  & \circIII &  &  & \circIII &  & $-$ \\
\gc 32 & \gc Chi et al.~\cite{chi2023privacy}            & \gc Protocol & \gc Blockchain & \gc ZKP & \gc & \gc \circIII &  \gc & \gc & \gc & \gc & \gc FP \\
33 & Khan et al.~\cite{app13031959}              & Protocol & Blockchain & ZKP &  & \circIII &  &  &  & \circIII & FP \\
\gc 34 & \gc Konasani~\cite{konasani2023decentralized}   & \gc Application/Domain & \gc Blockchain & \gc PKC & \gc \circIII & \gc \circIII & \gc & \gc \circIII & \gc & \gc & \gc $-$ \\
35 & Zeydan et al.~\cite{zeydan2023ssi}          & Application/Domain & Blockchain & PKC &  & \circIII &  &  &  & \circIII & BM \\
\gc 36 & \gc Kuznetsov et al.~\cite{kuznetsov2024enhanced} & \gc Protocol & \gc Blockchain & \gc ZKP & \gc & \gc \circIII & \gc & \gc & \gc & \gc & \gc FP \\
37 & Wang et al.~\cite{wang2024blockchain}       & System/Architecture & Blockchain & PKC &  & \circIII &  &  & \circIII &  & BM \\
\gc 38 & \gc Wang~\cite{wang2025automation}              & \gc Application/Domain & \gc Blockchain & \gc PKC & \gc & \gc & \gc \circIII & \gc & \gc \circIII & \gc & \gc BM \\
39 & Yun et al.~\cite{yun2025w3id}              & System/Architecture & Blockchain & ZKP &  & \circIII &  &  &  & \circIII & BM \\
\gc 40 & \gc Hauswirth et al.~\cite{hauswirth2003handling} & \gc Framework & \gc P2P & \gc PKC & \gc \circIII & \gc \circIII & \gc & \gc & \gc & \gc & \gc $-$ \\
41 & Bellavista et al.~\cite{bellavista2013peer} & System/Architecture & P2P & PKC &  & \circIII &  &  &  &  & BM \\
\gc 42 & \gc Brunner et al.~\cite{brunner2020did}        & \gc Protocol & \gc P2P & \gc PKC & \gc & \gc \circIII & \gc \circIII & \gc & \gc & \gc \circIII & \gc BM \\
43 & Stokkink et al.~\cite{stokkink2021truly}    & System/Architecture & P2P & PKC & \circIII & \circIII &  &  & \circIII & \circIII & BM \\
\gc 44 & \gc Hansen et al.~\cite{hansen2008privacy}       & \gc Framework & \gc Agnostic & \gc PKC & \gc & \gc & \gc \circIII & \gc \circIII & \gc & \gc & \gc US \\
45 & Davie et al.~\cite{davie2019trust}          & Framework & Agnostic & PKC &  & \circIII &  &  & \circIII &  & $-$ \\
\gc 46 & \gc Lux et al.~\cite{lux2020distributed}        & \gc System/Architecture & \gc Hybrid & \gc PKC & \gc & \gc \circIII & \gc \circIII & \gc & \gc & \gc \circIII & \gc BM \\
47 & Bambacht \& Pouwelse~\cite{bambacht2022web3} & Framework & Hybrid & PKC &  & \circIII &  &  & \circIII & \circIII & $-$ \\
\midrule
\multicolumn{5}{r}{\textbf{Total papers addressing each challenge:}} & \textbf{21} & \textbf{38} & \textbf{7} & \textbf{11} & \textbf{10} & \textbf{7} & \\
\bottomrule
\end{tabular}
\end{table*}

\begin{figure}
    \centering
    \includegraphics[width=0.65\linewidth]{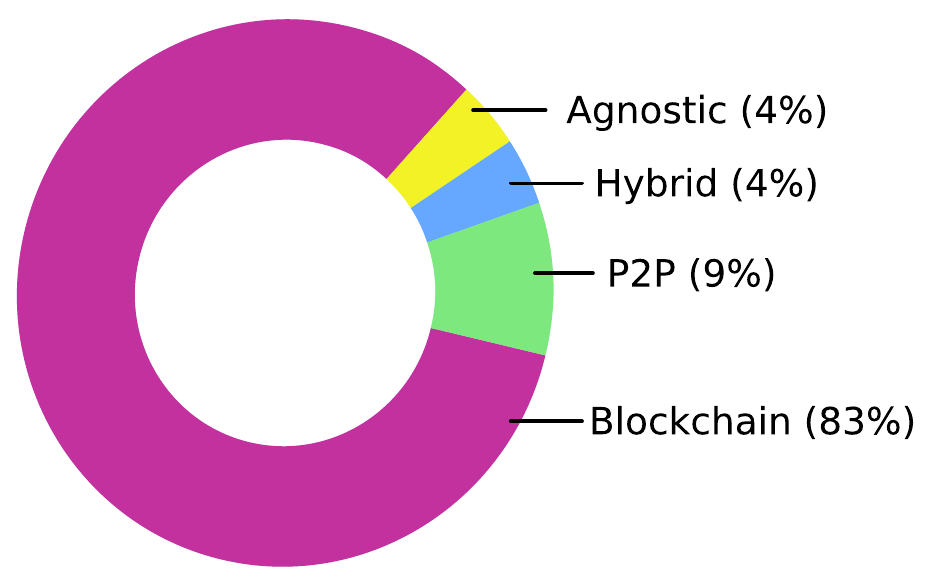}
    \caption{Targeted infrastructure. Our survey of scholarly work reveals the dominance of blockchain-focused work and very little focus on generalized targets.}
    \label{fig:infrastructure}
\end{figure}

\textbf{Trust Infrastructure.} The most striking finding is the overwhelming dominance of blockchain as the assumed trust infrastructure. Of the 47 papers analyzed, 39 (83\%) propose systems built on a specific blockchain platform, with Ethereum, Hyperledger Indy, or Hyperledger Fabric being the most common~\cite{liu2020blockchain, kuperberg2019blockchain, liu2017identity, bandara2021ssi, nusantoro2021blockchain, omar2018identity, dunphy2018first, naik2020self, cucko2021decentralized, maram2021candid}. Only 4 papers (9\%) explore purely peer-to-peer architectures without any blockchain dependency~\cite{stokkink2021truly, hauswirth2003handling, bellavista2013peer, brunner2020did}, and the remaining 4 (8\%) adopt hybrid or ledger-agnostic approaches~\cite{davie2019trust, bambacht2022web3, hansen2008privacy, lux2020distributed}. This blockchain bias means that much of the scientific literature is, in effect, studying \emph{blockchain-based identity management} rather than SSDI in its fullest sense. As shown in Fig.~\ref{fig:infrastructure}, this conflation of SSDI with blockchain is so pervasive that it may inadvertently discourage exploration of alternative trust anchors.

\begin{figure}[htbp]
    \centering
    \includegraphics[width=0.9\linewidth]{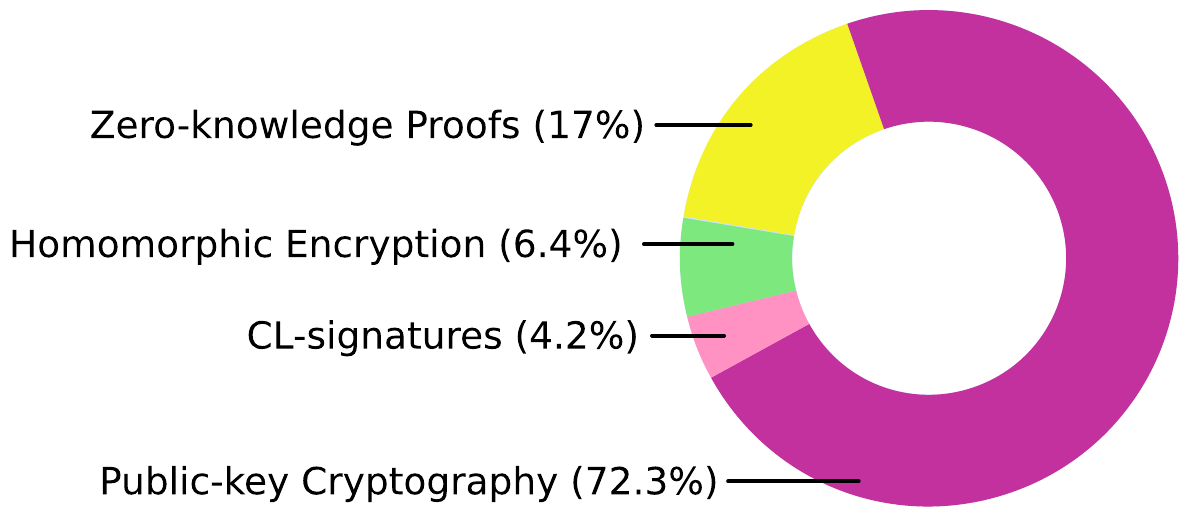}
    \caption{Cryptographic techniques in the scholarly literature. Public-key cryptography is the most popular type (72.3\%) followed by ZK proofs (17\%).}
    \label{fig:crypto}
\end{figure}

\textbf{Cryptographic Techniques.} The cryptographic landscape is dominated by standard public-key cryptography (RSA, ECDSA, EdDSA) for DID authentication. As shown in Fig.~\ref{fig:crypto}, a growing minority of papers incorporate advanced techniques: 8 papers (17\%) employ zero-knowledge proofs (ZKP) for selective disclosure or privacy-preserving authentication~\cite{yang2020zero, chi2023privacy, kuznetsov2024enhanced, app13031959, sarier2021efficient, maram2021candid, zhuang2022bcppt, yun2025w3id}, 3 papers (6\%) explore homomorphic encryption (HE) for privacy-preserving computation on identity attributes~\cite{shen2019efficient, chean2018authentication, drozdowski2019application}, and 2 papers (4\%) use CL-signatures for unlinkable credential presentations~\cite{stockburger2021blockchain, othman2018horcrux}. ZKPs are particularly promising for SSDI because they enable a user to prove possession of a credential or a derived property (e.g., ``I am over 18'') without revealing the credential itself. However, the computational cost and implementation complexity of ZKP systems remain significant barriers to practical deployment~\cite{vcapko2022state, lavin2024survey, sheybani2025zero}.

\begin{figure}[htbp]
    \centering
    \includegraphics[width=0.75\linewidth]{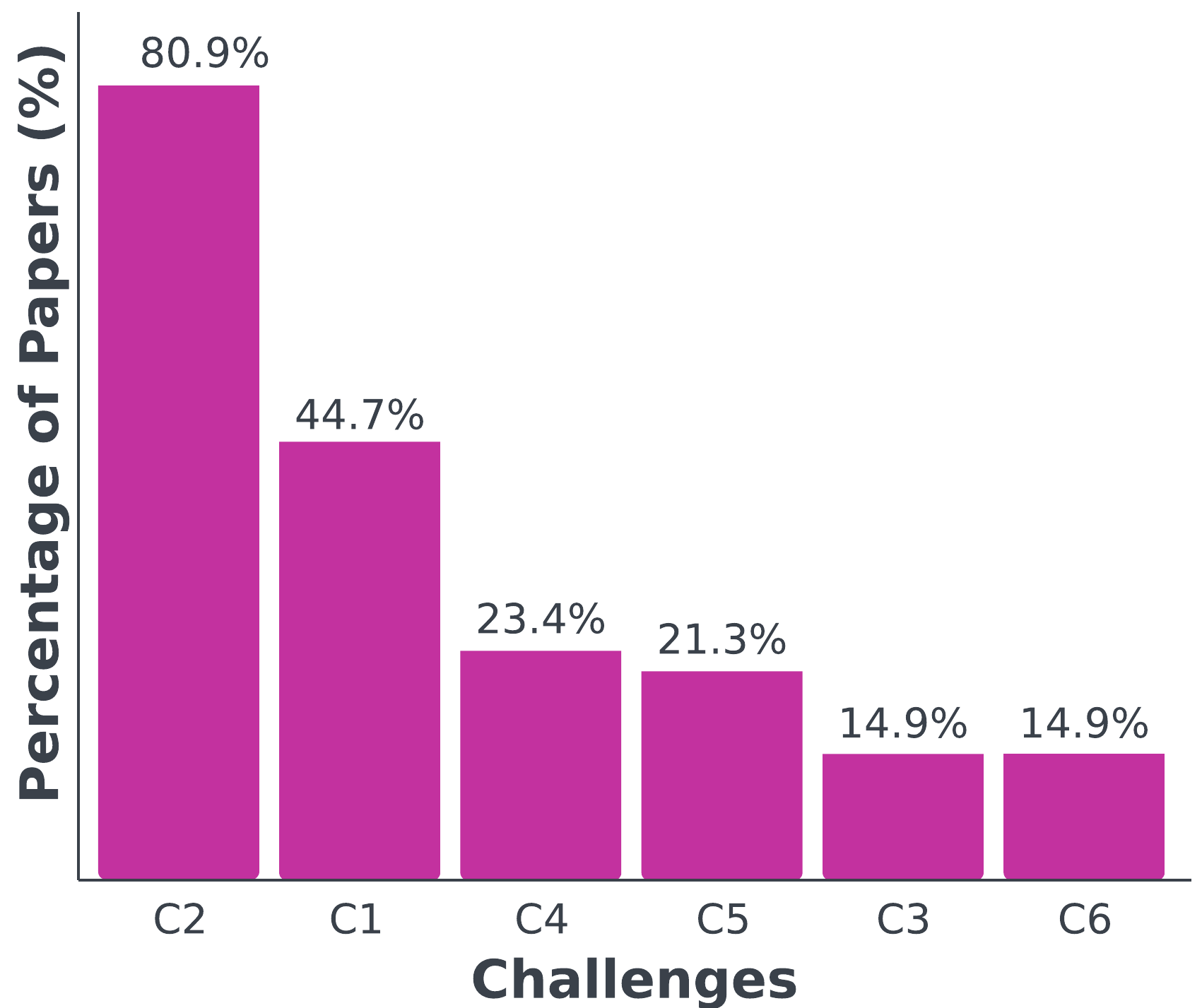}
    \caption{Percentage of scholarly work addressing particular SSDI challenges. The analysis reveals the predominance of focus on key management and protocols as well as identity binding, and a notable lack of focus on usability and meta-centralization.}
    \label{fig:total-papers-per-challenge}
\end{figure}

\begin{figure}[htbp]
    \centering
    \includegraphics[width=\linewidth]{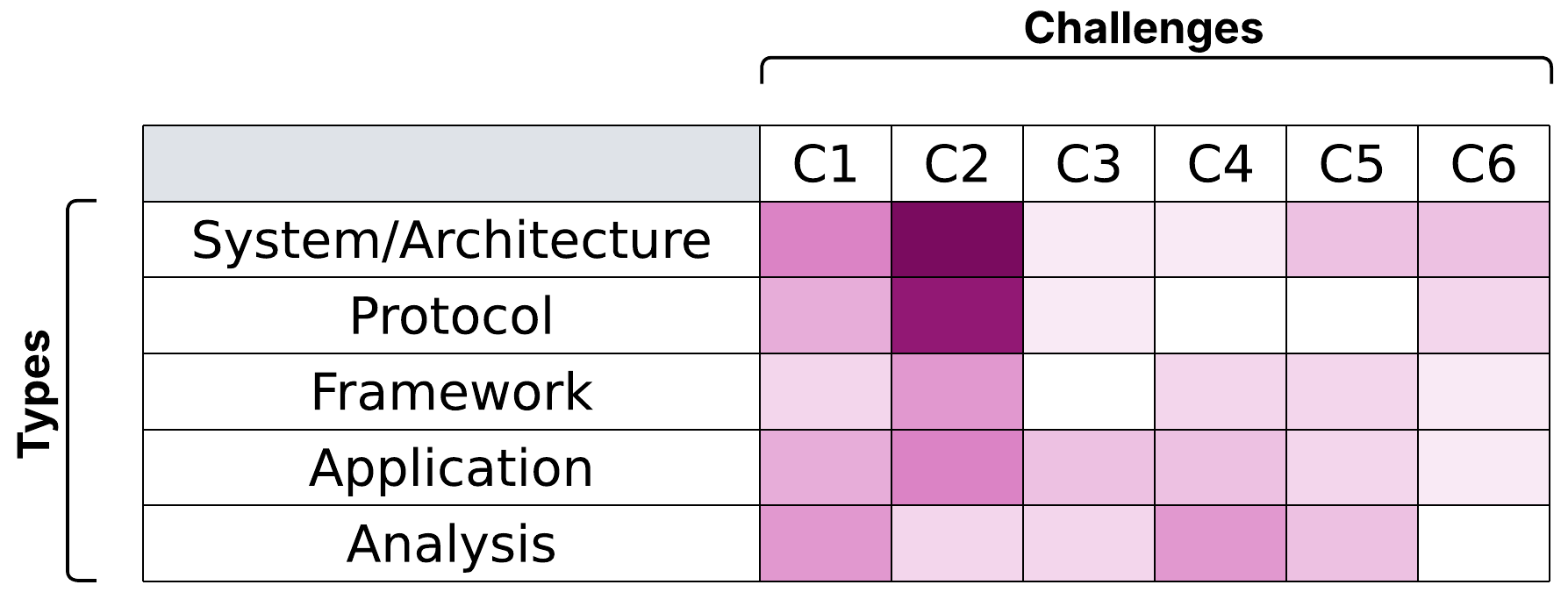}
    \caption{Heatmap of intersections between literature types and SSDI challenges. Systems, architectures, and protocols overwhelmingly target key management (C2), while usability (C3) and single infrastructure dependence (C6) receive minimal attention across all paper types.}
    \label{fig:heatmap}
\end{figure}

\textbf{Challenge Coverage.} Fig.~\ref{fig:total-papers-per-challenge} reveals how unevenly the six challenges are addressed. Challenge~2 (Protocols \& Key Management) is the most frequently targeted, appearing in 38 of 47 papers ($\approx$81\%), which is unsurprising, since proposing a new protocol or architecture inherently engages with this challenge. Challenge~1 (Identity Binding) is addressed by 21 papers (45\%), typically through credential-based or biometric mechanisms. In contrast, Challenge~3 (Usability) is engaged by only 7 papers (15\%), Challenge~4 (Oversight \& Regulation) by 11 (23\%), Challenge~5 (Critical-Mass Adoption) by 10 (21\%), and Challenge~6 (Single Infrastructure Dependence) by 7 (15\%). The neglect of usability is particularly concerning: as we argued in \S\ref{sec:usability}, usability is a first-order adoption barrier, yet the overwhelming majority of SSDI papers treat it as an afterthought or ignore it entirely (see Fig.~\ref{fig:heatmap}).

\textbf{Evaluation Methodology.} Rigorous evaluation is sparse. Only 7 of the 47 papers (15\%) include formal security proofs, and only 2 (4\%) report user studies or usability evaluations. The plurality (23 papers, 49\%) rely on implementation benchmarks (throughput, latency, gas cost), while 15 papers (32\%) offer only informal security arguments with no empirical evaluation. The near-absence of user studies is a critical gap: systems that are not evaluated with real users risk optimizing for the wrong objectives.

\begin{tcolorbox}[title=\textbf{Takeaways from Literature Analysis}, colback=white!95!black, colframe=gray!75!black, colbacktitle=gray!75!black, coltitle=white, boxsep=1.5pt, left=1mm, right=1mm, top=1mm, bottom=1mm]
Overall, the scientific literature reveals a field that is technically active but conceptually narrow. The dominance of blockchain and standard public-key cryptography, combined with the neglect of usability evaluation, reveals a community that has yet to grapple with the full breadth of challenges facing SSDI adoption. Usability, the challenge most likely to determine adoption success or failure, remains largely untested.
\end{tcolorbox}

\section{Real-World Applications}\label{sec:realworld}

To ground our analysis in practice, we identified 12 SSDI systems that have reached production deployment, pilot scale, or advanced development. Table~\ref{tab:realworld} summarizes our evaluation along five dimensions of self-sovereignty: user key control, permissionless issuance, credential portability, selective disclosure, and resistance to unilateral revocation.

\begin{table*}[]
\centering
\caption{\textbf{Real-world SSDI deployments evaluated along five dimensions of self-sovereignty.} \circIII = fully supported; \circII = partially supported; \circI = not supported, DEF = defunct, INA = inactive, ACD = active development, PAD = partial deployment, ACT = active/deployed.}
\label{tab:realworld}
\begin{tabular}{l|c|c|c|c|c|c}
\toprule
\multirow{2}{*}{\textbf{Project / Initiative}} & \textbf{Key} & 
\multirow{2}{*}{\textbf{Permissionless}} & \multirow{2}{*}{\textbf{Portability}} & \textbf{Selective} & \textbf{Revocation} & \textbf{Status} \\
 & \textbf{Control} & & & \textbf{Disclosure} & \textbf{Resistance} & \textbf{\textit{(as of Feb 2026)}} \\
\midrule
Sovrin~\cite{sovrin2018whitepaper,sovrin2019bcusecase,sovrin2019framework} & \circIII & \circII & \circIII & \circIII & \circII & DEF~\cite{sovrin_dissolved_2026} \\

\gc EU Digital Identity Wallet~\cite{eu_digital_identity_wallet_home_misc,eudigitalid_portal_misc} \gc & \gc \circIII & \gc \circI & \gc \circIII & \gc \circIII & \gc \circI & \gc ACD~\cite{eu_digital_identity_wallet_github} \\
British Columbia VON~\cite{von2019,bc_wallet_digital_gov_misc} & \circIII & \circII & \circII & \circIII & \circII & ACD, PAD~\cite{bcgov_bc_wallet_mobile} \\
\gc DIF ION~\cite{dif_ion_misc,dif_ion_repo_misc} & \gc \circIII & \gc \circIII & \gc \circII & \gc  \circI & \gc \circIII & \gc INA~\cite{dif_ion_commits} \\
SpruceID~\cite{spruceid_website_misc,spruceid_github_misc} & \circIII & \circIII & \circIII & \circII & \circIII & ACD~\cite{spruceid_github_repositories} \\
\gc uPort Veramo~\cite{veramo_website,veramo_github,uport_github} & \gc \circIII & \gc \circIII & \gc \circII & \gc \circII & \gc \circIII & \gc ACD~\cite{dif_veramo_commits_next,uport_project_repositories}\\
Jolocom~\cite{jolocom_project} & \circIII & \circII & \circII & \circIII & \circII & INA~\cite{jolocom_repositories} \\
\gc Trinsic~\cite{trinsic_website,trinsic_github} & \gc \circIII & \gc \circII & \gc \circIII & \gc \circIII & \gc \circII & \gc ACT~\cite{trinsic_status} \\
Truvera / Dock Labs / Dock.io~\cite{dock_website,docknetwork_github} & \circIII & \circII & \circIII & \circIII & \circII & ACT~\cite{truvera_status} \\
\gc KILT~\cite{kilt_whitepaper,kilt_github} & \gc \circIII & \gc \circIII & \gc \circII & \gc \circIII & \gc \circI & \gc DEF~\cite{kilt_website} \\
Ping Identity / ShoCard~\cite{pingidentity_website} & \circII & \circI & \circI & \circI & \circI & ACT~\cite{pingidentity_github} \\
\gc IDunion~\cite{idunion_github} & \gc \circIII & \gc \circII & \gc \circIII & \gc \circIII & \gc \circII & \gc DEF~\cite{idunion_secure_identity_management} \\
\bottomrule
\end{tabular}
\end{table*}

Several observations emerge from this evaluation. First, \emph{no deployed system achieves full self-sovereignty across all five dimensions}. Even the most decentralized systems make pragmatic compromises. For example, Sovrin's network provides strong selective disclosure via zero-knowledge proofs (using Camenisch-Lysyanskaya signatures) and gives users control over their keys, but the network itself is permissioned: only approved ``stewards'' can operate validator nodes~\cite{sovrin2018whitepaper, sovrin2019framework}. This means that while users have self-sovereign \emph{control}, they depend on a governed infrastructure for \emph{identity resolution}.

Second, government-backed initiatives such as the EU Digital Identity Wallet and British Columbia's Verified Organizations Network (VON)~\cite{von2019,sovrin2019bcusecase} achieve full support for key control and selective disclosure but lack permissionlessness and revocation resistance. This is by design: governments require the ability to revoke credentials (e.g., when a driver's license is suspended) and are unwilling to cede control over the issuance process. These systems are better described as \emph{user-centric} rather than fully \emph{self-sovereign}~\cite{josang2005user,ahn2009privacy,vossaert2013user}.

Third, commercially developed systems such as PingIdentity (which acquired ShoCard~\cite{pingidentity_shocard_2020}), despite positioning themselves within the SSDI ecosystem, exhibit significant centralization in practice. Users depend on the company's infrastructure for identity storage and verification, and credentials are not portable to competing platforms. These systems illustrate what might be called ``sovereignty washing''~\cite{uniqkey_sovereign_washing,civo_sovereignty_washing_2026,thenewstack_cloud_washing_2025}---adopting the SSDI label without delivering its core properties.

\begin{figure}[htbp]
    \centering
    \includegraphics[width=0.85\linewidth]{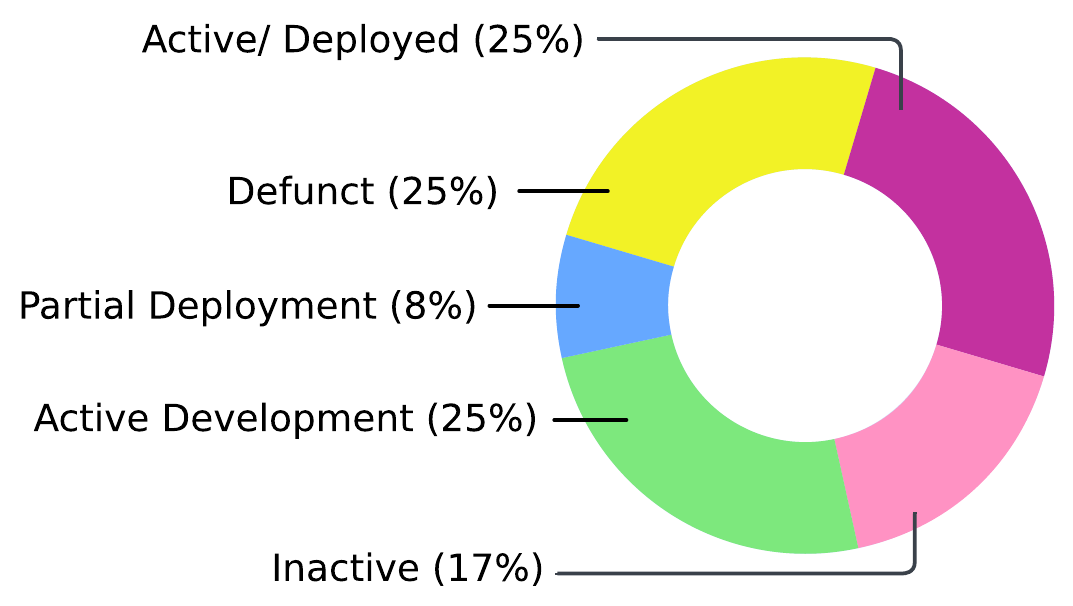}
    \caption{Status of real-world SSDI projects as of February 2026. Only one quarter of the projects are in active operation, and approximately 42\% are either defunct or inactive.}
    \label{fig:status}
\end{figure}

Fourth, the \emph{spectrum of self-sovereignty} that emerges from our evaluation suggests that practitioners should move beyond binary classifications. A more nuanced framework, akin to the NIST identity assurance levels~\cite{nist2020digital} but oriented toward sovereignty rather than assurance, would help stakeholders make informed decisions about which trade-offs are acceptable for a given context. We consider the five-dimension evaluation in Table~\ref{tab:realworld} a first step toward such a framework and identify its formalization as an important direction for future work.

Finally, as shown in Fig.~\ref{fig:status}, as of February 2026, only 3 of the 12 surveyed projects (25\%) were fully operational. Another 3 projects (25\%) were defunct and 2 (17\%) were effectively inactive\footnote{Based on the assessment of their code repositories and/or their websites.}, meaning that approximately 42\% of the ecosystem has stalled. The remaining 4 projects (33\%) were either under active development or in partial deployment.

Our evaluation of real-world deployments underscores a broader theme: the gap between the SSDI ideal and practical deployment is not merely technical but reflects deep tensions between user autonomy, institutional accountability, and regulatory compliance.

\begin{tcolorbox}[title=\textbf{Takeaways from Assessing Real-World Applications}, colback=white!95!black, colframe=gray!75!black, colbacktitle=gray!75!black, coltitle=white, boxsep=1.5pt, left=1mm, right=1mm, top=1mm, bottom=1mm]
Very few real-world SSDI deployments are ready for use. Moreover, those that are actively deployed do not meet many SSDI criteria, supporting the concern that the area is affected by ``sovereignty washing.''
\end{tcolorbox}

\section{Frontiers of SSDI}\label{sec:frontiers}
Drawing on our challenge taxonomy, literature analysis, and real-world evaluation, we identify five frontier areas that are poised to shape the future of SSDI research and practice. These frontiers do not map one-to-one to the six challenges identified in \S\ref{sec:challenges}; rather, each frontier cross-cuts multiple challenges, and we indicate these connections where relevant.

\subsection{Frontier 1: Privacy-Enhancing Cryptography}
Zero-knowledge proofs (ZKPs) and homomorphic encryption (HE) represent the most promising cryptographic tools for advancing SSDI. ZKPs enable a prover to convince a verifier of a statement's truth without revealing any information beyond the statement itself~\cite{sun2021survey, chi2023privacy, gupta2025zero}. In the SSDI context, ZKPs can enable powerful privacy-preserving interactions: proving that one is over a certain age without revealing one's date of birth, proving membership in an organization without revealing one's role, or proving solvency without disclosing account balances. Recent advances in ZKP efficiency, particularly zk-SNARKs, zk-STARKs, and Plonk-based systems~\cite{lavin2024survey, sheybani2025zero, xing2025zero}, are bringing these capabilities closer to practical deployment. However, significant work remains in making ZKP-based credentials interoperable across DID methods and wallet implementations.

Homomorphic encryption allows computation on encrypted data without decryption~\cite{acar2018survey, armknecht2015guide, frederick2015homomorphic}, enabling scenarios where a verifier can evaluate a predicate over a user's encrypted attributes. While fully homomorphic encryption remains computationally expensive, partially homomorphic schemes may be sufficient for many identity verification tasks~\cite{drozdowski2019application, chen2023quantum}. Key open questions at the intersection of ZKPs, HE, and SSDI include how to standardize proof formats across wallet implementations and how to manage the trust assumptions that different ZKP constructions introduce.

\subsection{Frontier 2: Web3 Integration}
The broader Web3 movement, encompassing decentralized finance (DeFi), non-fungible tokens (NFTs), and decentralized autonomous organizations (DAOs), provides both a natural application domain and a source of infrastructure for SSDI~\cite{bambacht2022web3, liu2023web3, bashir2023mastering}. DIDs and verifiable credentials can serve as the identity layer for Web3 applications, enabling Sybil-resistant governance in DAOs, KYC-compliant DeFi~\cite{konasani2023decentralized}, and provenance-tracked digital assets. Conversely, Web3 infrastructure (smart contracts, decentralized storage via IPFS or Storj~\cite{storj2018v3}, cross-chain bridges) can provide the technical substrate for SSDI systems~\cite{wang2024blockchain, yun2025w3id}. The challenge lies in achieving this integration without tethering SSDI to the volatility and regulatory uncertainty of the cryptocurrency ecosystem. Furthermore, the current reliance of Web3 on the blockchain ecosystem means that Web3 systems suffer, by proxy, from the multidimensional scalability issues~\cite{ivanov2021blockumulus} inherent to blockchain.

\subsection{Frontier 3: IoT and Machine Identity}
The Internet of Things introduces a new dimension to SSDI: identity for devices and autonomous systems~\cite{mahalle2022identity, bouras2021lightweight, venkatraman2022developing, omar2018identity}. A smart building sensor, an autonomous vehicle, or a supply-chain tracking device each needs a verifiable identity for secure communication and data provenance~\cite{wang2025automation, zeydan2023ssi}. The challenges are distinct from human identity: devices cannot perform cognitive tasks like reviewing disclosure requests, key recovery must be automated, and the sheer scale (billions of devices) demands lightweight protocols. Self-sovereign \emph{machine} identity is an emerging research area that inherits all six challenges identified in \S\ref{sec:challenges} while introducing additional ones specific to resource-constrained, autonomous agents, such as automated key lifecycle management, constrained-device attestation, and scalability to billions of endpoints~\cite{wooldridge2009introduction, ivanov2022autothing}.

\subsection{Frontier 4: Regulatory Harmonization}
The regulatory landscape for digital identity is evolving rapidly but remains fragmented across jurisdictions~\cite{oecd2021digital, Brennen2024The}. The EU's eIDAS~2.0 regulation, India's Aadhaar system (a centralized biometric approach that SSDI seeks to supplant), Canada's Pan-Canadian Trust Framework~\cite{diacc2019pctf}, and various U.S.\ state-level initiatives each embody different philosophies about the balance between individual sovereignty and state oversight. Harmonizing these approaches (or at least establishing mutual recognition frameworks) is essential for SSDI to function across borders. The World Economic Forum's blueprint for digital identity~\cite{wef2018blueprint} and the OECD's guidelines~\cite{oecd2021digital} provide starting points, but translating high-level principles into interoperable legal and technical standards remains an open challenge.

\subsection{Frontier 5: Usability-First Design}
Perhaps the most impactful frontier is a paradigm shift in how SSDI systems are designed: from \emph{technology-first} (building the cryptographic and protocol infrastructure, then adding a user interface) to \emph{usability-first} (starting with user needs and cognitive models, then selecting technologies that fit). The field of human-computer interaction has long argued that security technologies fail when they are designed without the user in mind~\cite{borcea2006what}. SSDI is no exception. Novel interaction paradigms, such as ambient or implicit authentication that reduce the cognitive burden on users while preserving the privacy benefits of self-sovereignty, represent a particularly promising direction. More broadly, future work should invest heavily in user research (longitudinal studies, field deployments, A/B testing of wallet designs) and develop standardized usability benchmarks for SSDI systems.

\paragraph{Additional emerging themes.}
Beyond these five frontiers, several other themes deserve attention. The \emph{economics of SSDI}---who pays for the infrastructure, and how are incentives aligned across issuers, holders, and verifiers---remains undertheorized. The \emph{intersection of SSDI with artificial intelligence}, including the use of AI for identity verification, the risk of deepfakes undermining biometric binding, and the need for machine-readable credentials in AI-driven workflows, is a rapidly emerging concern~\cite{jung2023post, jeon2023use}. Finally, the \emph{social and ethical dimensions} of self-sovereignty, particularly for vulnerable populations who may lack the technical literacy, device access, or social networks required to manage self-sovereign credentials, demand sustained attention from researchers, policymakers, and civil society organizations~\cite{whitley2021rethinking, wef2021digital}.

\begin{tcolorbox}[title=\textbf{Looking Ahead}, colback=white!95!black, colframe=gray!75!black, colbacktitle=gray!75!black, coltitle=white, boxsep=1.5pt, left=1mm, right=1mm, top=1mm, bottom=1mm]
Self-Sovereign Digital Identities face six major challenges that current research addresses only partially and in isolation, necessitating more comprehensive work across five critical frontiers: privacy-enhancing cryptography, Web3 integration, IoT and machine identity, regulatory harmonization, and usability-first design.
\end{tcolorbox}

\section{Conclusion}\label{sec:conclusion}
Self-Sovereign Digital Identity represents a compelling vision for the future of digital identity management, one in which individuals and organizations control their own identity assertions, share only what is necessary, and depend on no single intermediary. This paper has presented a comprehensive systematization of knowledge on SSDI, organized around six challenges that collectively explain why this vision has not yet been realized in practice: the difficulty of binding digital identities to real-world entities, the immaturity of key management and communication protocols, significant usability barriers, regulatory and oversight gaps, the bootstrapping problem of achieving critical-mass adoption, and the systemic risk of depending on a single decentralized infrastructure.

Our analysis of 47 scientific publications reveals that the research community has equated SSDI with blockchain to a degree that limits the design space and leaves alternative architectures underexplored. Our evaluation of 12 real-world SSDI systems demonstrates that self-sovereignty is best understood as a spectrum, with no current system achieving full sovereignty across all dimensions. Government-backed initiatives prioritize institutional control over user autonomy; some commercially developed systems risk ``sovereignty washing''; and even the most decentralized projects make pragmatic compromises that fall short of the self-sovereign ideal.

Looking ahead, we identify five frontier areas that are poised to shape the field: privacy-enhancing cryptography (particularly zero-knowledge proofs and homomorphic encryption), Web3 integration, IoT and machine identity, regulatory harmonization across jurisdictions, and a fundamental shift toward usability-first design. Progress on these frontiers will require collaboration across disciplines (cryptography, systems security, human-computer interaction, law, economics, and public policy) and a willingness to move beyond the blockchain-centric paradigm that currently dominates the literature.

We hope that this systematization will serve as a reference point for researchers entering the field, a guide for practitioners evaluating SSDI technologies, and a call to action for the community to address the challenges we have identified. The evolutionary shift from centralized to federated to self-sovereign digital identities is not inevitable; it must be earned through rigorous research, thoughtful design, and sustained engagement with the societal implications of identity technology.

\section{Acknowledgement}
This material is based upon work supported by the National Science Foundation under Grant No. 2348344. Any opinions, findings, and conclusions or recommendations expressed in this material are those of the author(s) and do not necessarily reflect the views of the National Science Foundation.


\bibliographystyle{ACM-Reference-Format}
\bibliography{bibliography}

\end{document}